# Algorithm to Compute Squares of 1st "N" Natural Numbers Without Using Multiplication.

-Rajat Tandon

Computer Science & Engineering, Manipal Institute of Technology, Manipal, Karnataka - India

rajattandonmit@gmail.com

## ABSTRACT

Processors may find some elementary operations to be faster than the others. Although an operation may be conceptually as simple as some other operation, the processing speeds of the two can vary. A clever programmer will always try to choose the faster instructions for the job. This paper presents an algorithm to display squares of $1^{st}$ "N" natural numbers without using multiplication ('*' operator). Instead, the same work can be done using addition ('+' operator). The results can also be used to compute the sum of those squares. If we compare the normal method of computing the squares of $1^{st}$ "N" natural numbers with this method, we can conclude that the algorithm discussed in the paper is more optimized in terms of time complexity.

## 1. KEYWORDS

Time complexity, Square

## 2. INTRODUCTION

Time complexity is defined as a measure of the amount of time taken to execute an algorithm. It is commonly estimated by counting the number of elementary operations performed by the algorithm, where an elementary operation takes a fixed amount of time to perform. Addition requires lesser time than multiplication. If the value of "N" is small, then there will not be much time difference. However, for a large value of "N", the difference cannot be ignored. Table 1 shows some of the elementary operations and their Time complexities.

## 3. ALGORITHM

Algorithm Square_of_first_n_nos(n)
//Input: Any positive integer "n"
//Output: Squares of $1^{st}$ "n" natural numbers

a ← 1
square ← 0

for i ← 1 to n do
    square ← square + a
    a ← a + 2

## 4. EXPLAINATION

The logic behind the algorithm is that all perfect squares of natural numbers differ by an odd number. For example, if we start from 1 and add 3 to it, we will obtain the square of 2. Similarly, if we add 5 to the previous value of the variable "square" then we get 9 which is the square of 3.

## 5. OUTPUT TRACING

Initially the values of the variables are shown in Table 2.

| a | square |
|---|--------|
| 1 | 0 |

Table 2

Let us consider the first iteration. The value of the variable "square" is added with the variable "a" to get the square of the first number. Then the value of the variable "a" is incremented by 2 in order to compute the square of the next number in the next iteration.

| OPERATION | INPUT | OUTPUT | ALGORITHM | COMPLEXITY |
|-----------|-------|--------|-----------|------------|
| Addition | Two n-digit numbers | One n+1 – digit number | Schoolbook addition with carry | $\Theta(n)$ |
| Subtraction | Two n-digit numbers | One n+1 – digit number | Schoolbook subtraction with borrow | $\Theta(n)$ |
| Multiplication | Two n-digit numbers | One 2n-digit number | Schoolbook long multiplication | $O(n^2)$ |
| Division | Two n-digit numbers | One n-digit number | Schoolbook long division | $O(n^2)$ |

Table 1: Elementary operations with their Time complexities

The values of the variables after the first iteration are shown in Table 3.

| a | square |
|---|---|
| 3 | 1 |

Table 3

Let us consider the second iteration. The value of the variable "square" is added with the variable "a" to get the square of the first number. Then the value of the variable "a" is incremented by 2 in order to compute the square of the next number in the next iteration.

The values of the variables after the second iteration are shown in Table 4.

| a | square |
|---|---|
| 5 | 4 |

Table 3

Similarly, the loop runs till the value of variable "i" reaches "n".

## 6. ADDITIONAL RESULT

With the aid of the logic used in the algorithm, we can derive a result stated below:

Square(n) = Square(n-1) + 2*n - 1
Or
Square(n+1) = Square(n) + 2*(n+1) -1

The above result states that, we can find out the square of a natural number given the square of its preceding natural number.

## 7. CONCLUSIONS

The algorithm discussed in the paper presents a more optimized solution of computing the squares of $1^{st}$ "N" natural numbers in terms of Time complexity, than the normal algorithm. Usually, addition and subtraction are preferred to multiplication and division, because the latter operations require more overhead if Time complexity issue is involved. The same can also be used to compute the sum of squares of $1^{st}$ "N" natural numbers.